# Quantum dot single photon source on SiN integrated with coupled crossover waveguides

Akinari Fujita[1]*, Hironobu Yoshimi[1], Natthajuks Pholsen[2], Masahiro Kakuda[3], Makoto Okano[4], Satoshi Iwamoto[2,5], Yasuhiko Arakawa[3], and Yasutomo Ota[1]*

[1]*Department of Applied Physics and Physico-Informatics, Keio University, Yokohama, Kanagawa 223-8522, Japan*

[2] *Institute of Industrial Science, The University of Tokyo, Meguro, Tokyo 153-8505, Japan*

[3]*Institute for Nano Quantum Information Electronics, The University of Tokyo, 4-6-1 Komaba, Meguro-ku, Tokyo, Japan*

[4]*National Institute of Advanced Industrial Science and Technology, Tsukuba, Ibaraki 305-8569, Japan*

[5]*Research Center for Advanced Science and Technology, The University of Tokyo, Meguro, Tokyo 153-8904, Japan*

*E-mail: fujitaakinari@keio.jp; ota@appi.keio.ac.jp

Hybrid integration of InAs/GaAs quantum dot (QD) single-photon sources (SPSs) is a promising approach for introducing quantum light into SiN photonic integrated circuits. However, the large refractive-index mismatch between GaAs and SiN poses a challenge for efficient optical coupling. Here, we propose and experimentally demonstrate hybrid integration of an InAs/GaAs QD-SPS on SiN using a coupled crossover waveguide structure. A photonic crystal nanocavity is employed for coupling QD emission into a GaAs waveguide, which efficiently transfers photons to a SiN waveguide at the crossover section. We observed Purcell-enhanced single-photon emission, on-chip propagation, and outcoupling through a SiN grating coupler.

Template for APEX (May 2024)SiN photonic integrated circuitry has gain attention as a promising platform for large-scale photonic quantum information processing owing to its ultra-low-loss waveguiding, broadband transparency, and compatibility with CMOS manufacturing technology [1-3]. However, the integration of high-performance single photon sources (SPSs) remains a significant challenge despite recent progress in the integration [4,5] and the creation of quantum emitters in SiN [6-8]. To address this issue, hybrid integration approaches are regarded as straightforward alternatives because they enable the incorporation of high-quality SPSs developed in different material platforms while maintaining compatibility with CMOS processes [9,10].

Among various solid-state SPSs capable of on-demand operation, InAs/GaAs quantum dots (QDs) stand out for their ability to simultaneously achieve high brightness, coherence, and purity [11]. Hybrid integration of QD-SPSs into SiN circuitry has been examined most notably using pick-and-place assembly techniques such as micromanipulation and transfer printing [12-15]. While the placement of InAs/GaAs QD-SPSs on SiN circuitry has been successfully demonstrated, efficient optical coupling is largely hindered by the large refractive index difference between GaAs ($n = 3.4$) and stoichiometric $Si_3N_4$ ($n = 2$). A major coupling strategy employs tapered waveguides; however, the strong refractive-index mismatch requires long tapers terminated with sharp tips to achieve near unity efficiency [5,12,13,15-18]. Such structures are vulnerable to position and rotational misalignments, thereby hampering high-efficiency operation in experiment. Another approach employs photonic crystal (PhC) nanobeam cavities and has demonstrated an estimated optical coupling efficiency exceeding 50% [14]. However, in this scheme, the nanobeam cavity is designed to reduce its effective refractive index to attain phase matching with a SiN waveguide, at the expense of tight light confinement beneficial for enhancing light matter interactions. The use of two-dimensional (2D) PhC nanocavity is even more difficult for direct evanescent coupling to SiN waveguides due to their larger effective indices than those of the 1D structures.

Recently, coupled crossover waveguides, which consist of vertically separated linear crossing waveguides evanescently interacting each other, have emerged as a novel scheme for transferring photons between waveguides [19, 20]. With proper design, these structures can support near unity coupling between waveguides made from different materials without requiring long coupling sections. In addition, their linear crossover geometry makes them

Template for APEX (May 2024)tolerant to in-plane positional misalignment. Nevertheless, this approach has remained largely unexplored, in particular for material pairs with large refractive-index mismatch, such as GaAs and SiN.

In this work, we propose and experimentally demonstrate hybrid integration of an InAs/GaAs QD-SPS on a SiN photonic integrated circuit using a coupled crossover waveguide structure. We employed a 2D PhC nanocavity for funneling QD emission into a GaAs waveguide [21, 22], which couples to a single-mode SiN waveguide at the linear crossover region. We numerically designed the entire system to support a coupling efficiency exceeding 90% for QD emission into the SiN waveguide. Experimentally, we fabricated the designed structure using transfer printing and observed Purcell-enhanced single photon generation, on-chip waveguiding, and outcoupling through a SiN grating coupler.

Figure 1(a) shows a schematic of the proposed device structure. An InAs/GaAs QD acting as a SPS is embedded in a 2D PhC nanocavity, which funnels QD emission into a single cavity mode. The nanocavity is coupled to a W1 PhC waveguide [23], which is subsequently connected to a GaAs wire waveguide. Photons propagating in the GaAs waveguide are transferred to an underlying SiN waveguide at the linear crossover region via evanescent coupling [19]. The entire structure is covered by a glass thin film, which suppresses optical loss in the nanocavity by suppressing unwanted polarization conversion due to vertical asymmetry [24]. In Fig. 1(a), $\beta$ represents the coupling efficiency of QD emission into the cavity mode. $\eta_1$ and $\eta_2$ quantify the connectivity between the cavity and the W1 waveguide, and between the W1 and GaAs wire waveguides, respectively. $\eta_3$ denotes the coupling efficiency at the coupled crossover waveguide structure. We assume lossless light propagation outside the key components; the taper section embedded in the GaAs waveguide is also assumed to be lossless. In this system, the total efficiency is expressed as $\beta\eta_1\eta_2\eta_3$.

First, we designed a nanocavity coupled to the W1 waveguide to achieve high $\beta$ and $\eta_1$. We considered a multistep heterostructure PhC nanocavity [25] located in a W0.9 waveguide as shown in Fig. 1(b). The W0.9 waveguide is defined by removing airholes in a triangular lattice 2D PhC along the ΓK direction and by shrinking the width of the generated waveguide by 10%. The lattice constant, air-hole radius, and slab thickness are set to $a$ = 240 nm, $r$ = 0.22$a$, and $h$ = 130 nm, respectively. To form a nanocavity, the PhC lattice in the heterostructure region was elongated by 0.96% and 0.48% along the ΓK direction. On top of the structure, a 300-nm-thick glass thin film is placed to partially recover refractive index symmetry in the normal direction. Without forming the coupling W1 waveguide, the structure supports a fundamental cavity mode resonating at $\lambda$ = 940 nm with an intrinsic quality factor of $Q = 2.5 \times 10^6$ and a mode volume of $1.83(\lambda/n)^3$. The field distribution of the cavity mode calculated by the 3D finite difference time domain (FDTD) method is shown in the right panel of Fig. 1(b). Note that the intrinsic Q factor degrades to $Q = 7.9 \times 10^4$ in the absence of the glass thin film.

Now, a W1 waveguide is introduced near the cavity region [23], as shown in Fig. 1(c). The waveguide position and separation from the cavity are carefully optimized so that nearly all light leakage occurs into the W1 waveguide, resulting in a high $\eta_1$. Under the presence of the W1 waveguide, the cavity mode exhibits a waveguide-limited Q factor of $Q$ =1,020, corresponding to a Purcell enhancement factor of 42 for a QD located at the cavity field maximum. Given that the strong photonic bandgap effect in the 2D PhC suppresses non-cavity mode spontaneous emission by a factor of 10 [26], the maximum $\beta$ is estimated to reach 99.8%. Furthermore, from the contrast between the intrinsic and waveguide-coupled Q factors, $\eta_1$ is estimated to exceed 99.9%. The efficient cavity-waveguide coupling can be confirmed by the field distribution shown in the right panel of Fig. 1(c).

Then, we optimized $\eta_2$ for efficient connection between the W1 and GaAs waveguides. We employed a tapered PhC waveguide as shown in Fig. 1(d) [27]. We linearly expanded the PhC lattice away from the waveguide core over the length of $10a$. At the end of the taper, the PhC waveguide is expanded by $4r$ and is connected to a GaAs wire waveguide with a width of $\sqrt{3}a + 4r$. In this configuration, a high $\eta_2$ of 92.7% was computed by the 3D FDTD method, as confirmed in the field distribution in Fig. 1(d). We consider that further optimization of the hole positions will significantly improve the transmission efficiency [28].

Now, we design a coupled crossover waveguide structure for transferring photons into the SiN waveguide. Figure 2(a) shows top and cross-sectional structural schematics. The top GaAs waveguide evanescently interacts with the SiN waveguide underneath. The system in essence behaves as a directional coupler and therefore requires phase matching between the GaAs and SiN waveguides. The coupling strength can be controlled by the gap distance $g$ between the two layers [19]. We designed a single mode SiN waveguide at the wavelength of the operation by setting its height and width to 350 nm and 500 nm, respectively. To match the effective indices, the width of the GaAs waveguide is assumed to shrink to 190 nm using the taper section. The GaAs waveguide now appears narrow but leaves a sufficient margin of manufacturability with electron beam lithography used for patterning the QD-SPSs. Figure 2(b) shows field distributions around the linear crossover structure designed with a crossing angle of 8° and $g = 200$ nm. The upper-left panel shows the field in the GaAs layer, showing the propagation of the incident light. After passing through the crossover, light is transferred to the SiN layer as depicted in the lower-left panel. The right panels in Fig. 2(b) show cross-sectional field profiles recorded at the slice positions indicated in Fig. 2(a). The light transfer from the upper to lower layer is clearly visible. At the wavelength of 940 nm, the coupling efficiency $\eta_3$ is deduced to be a near unity value of 99.4%. Figure 2(c) summarizes calculated coupling efficiencies as a function of crossing angle for three different $g$. In each case, there is an optimal angle supporting near unity $\eta_3$. This observation can be interpreted from an analogy with directional couplers, in which the optimal coupling length is inversely proportional to the coupling strength. In the current structure, the gap and crossing angle respectively determine the interaction strength and length between the two waveguides. Importantly, thanks to the short interaction length of only ~ 5 μm at a relatively large crossing angle of 8°, the dependence of $\eta_3$ on the crossing angle is relatively weak, relaxing the required accuracy of the integration process. We note that $\eta_3$ also only mildly depends on the wavelength of operation and the wavelength range supporting >90% efficiency spans over 30 nm. Integrating all the design information, we deduce a peak total coupling efficiency $\beta\eta_1\eta_2\eta_3$ of QD radiation into the SiN waveguide to be 91.8%, which is predominantly limited by the relatively-low $\eta_2$ value in the current design.

The fabrication of the designed SPS structure starts from the independent preparation of the optical elements. First, a QD wafer was grown by molecular beam epitaxy on a GaAs substrate. A layer of InAs/GaAs QDs was embedded in a 130-nm-thick GaAs slab on top of an AlGaAs sacrificial layer. In the GaAs slab, we patterned the designed 2D-PhC nanocavities connected to W1 and wire waveguides by electron beam lithography followed by dry and wet etching processes. The whole structure is enclosed in a supporting frame, making it transferrable at once. Figure 3(a) shows a scanning electron microscope (SEM) image of a fabricated SPS structure after undercutting the sacrificial layer. In the device, the GaAs wire waveguide incorporates a 6-μm-long adiabatic taper converting the waveguide width from 640 nm to 190 nm. At the end of the waveguide, a grating coupler is introduced to monitor photons in the GaAs wire waveguide. We note that, in the PhC nanocavity used in experiment, a double-periodic 3% modulation of airhole radii was introduced around the Template for APEX (May 2024)cavity center to enhance vertical radiation for auxiliary optical measurements [29]. In addition, *a* is set to 245 nm to tune the resonance wavelength, and the lattice modulation strengths to form the heterostructure cavity region were modified to 2.88% and 1.44%. These modifications slightly reduce $\eta_1$ to 99.7% and do not noticeably change $\beta$ in design.

In parallel, we prepared transferable 300 nm-thick glass thin films on a Si substrate. The glass layer was first deposited by chemical vapor deposition and was patterned by electron beam lithography followed by dry etching. The air-bridge structure was formed by dissolving Si below $SiO_2$ by a TMAH solution [21,30]. An optical microscope image of a fabricated glass thin film is shown in Fig. 3(b). Meanwhile, we obtained SiN waveguides from a foundry service. A bird's eye SEM image of a cleaved SiN waveguide is displayed in the upper panel of Fig. 3(c). The SiN waveguide chips are planarized by a spin-on-glass material. The clad thickness above SiN was controlled to be 150 nm as shown in the bottom panel of Fig. 3(c). The SiN waveguides are also terminated with grating couplers to radiate photons into free space.

Finally, we assemble the prepared photonic components to build the designed on-SiN QD-SPS using transfer printing [31]. First, a piece of the GaAs structure was placed onto a SiN waveguide. We aligned the position and rotation angle of the GaAs wire waveguide with respect to the SiN waveguide under an optical microscope equipped in our transfer-printing apparatus, which routinely supports a rotational accuracy of approximately 0.1°. The length of the coupling section in the SiN waveguide is 10 μm, which is sufficiently long for overlaying it with the GaAs waveguide at a rotation angle of 9°. Here, we modified the crossing angle from the designed value of 8° to increase $\eta_3$ in the experiment, considering that the geometries of the fabricated components deviate from the designed values. Then, a glass thin film was placed over the entire structure. This process step requires special care as air bubbles often remain beneath the glass thin film. To remove them, we scrubbed the glass by the transfer stamp or repeated the glass transfer itself after delaminating the transferred glass thin film. Figure 3(d) shows an optical microscope image of a completed device, showing the successful assembly. The crossover region is homogeneously formed, judging from the constant color. We note that the small fragment of the GaAs membrane in the picture was induced during the repeated transfer processes to remove air bubbles.

The fabricated device was optically characterized with a low-temperature micro-photoluminescence (PL) setup. Our setup employs an objective lens with a numerical aperture of 0.4, which was used for focusing pump laser light, collecting PL signals and imaging the sample. PL signals were analyzed with a spectrometer, together with a CCD camera for recording spectra or superconducting single photon detectors for time-resoled PL spectroscopy and intensity correlation measurements. In the latter measurements, the spectrometer was used as a spectral bandpass filter. Figure 4(a) shows PL spectra measured at 55 K, recorded above the cavity or through the GaAs or SiN grating out-couplers under 633-nm continuous-wave (CW) excitation focused at the cavity center. The pump power was 2 µW. The two spectra measured through the out couplers well resemble each other, showing a cavity peak at 917.9 nm and a bright QD peak at 918.3 nm, while signals away from the cavity resonance are weak. In contrast, the spectrum measured above the cavity shows a largely-suppressed cavity peak with many QD peaks outside the cavity resonance. We note that the QD peak near 918.3 nm is accidentally from a different QD, confirmed via polarization resolved PL measurements. The comparison indicates that only the cavity-coupled signal can be efficiently connected to the waveguides. From multi-peak fitting to the spectrum from the SiN grating, we deduced a cavity Q factor of 2,070, which is lower than the average Q-factor of 3,600 measured for the cavities that do not have adjacent W1 waveguides. From the observed reduction in Q factor, we estimated $\eta_1$ to be 42.6%, which is predominantly limited experimentally by the significantly low intrinsic Q-factor of 3,600. The PL count difference between the spectra taken from the GaAs and SiN gratings hints the coupling efficiency $\eta_3$ of the coupled crossover waveguide structure. From 3D FDTD simulations, the efficiencies of the GaAs and SiN out-couplers were estimated to be 4.6% for both. After taking these efficiencies into account, we deduced $\eta_3$ to be 43.0%. The non-ideal $\eta_3$ value is likely a result of fabrication imperfections, such as waveguide bending or geometrical deviations from the designed values. We note that the influence of the air bubbles between the glass film and the chip appears to be negligible, as they do not overlap with the crossover region.

Figure 4(b) shows a plot of temperature-dependent PL spectra recorded through the SiN grating coupler. The QD peak is enhanced around the QD-cavity resonance, suggesting the presence of Purcell enhancement and the efficient funneling of QD radiation into the cavity and thus a high $\beta$. We then performed time-resolved PL spectroscopy at 50K for the QD peak at 918.5 nm under pulsed excitation with an average pump power of 2 µW using a pulsed laser oscillating at 750 nm. The recorded PL decay, together with that for bulk QDs, is shown in Fig. 4(c). From fitting, the QD lifetime under the cavity resonance was deduced to be 0.2 ns, which is three times faster than that for the bulk QDs, yielding a Purcell factor of 3.3. We estimated $\beta$ to be 97.1% after accounting for the photonic bandgap effect. Combining the experimentally estimated coupling efficiencies of $\beta$, $\eta_1$, and $\eta_3$, together with the numerically estimated $\eta_2$, the overall efficiency of the fabricated on-SiN QD-SPS is estimated to be 16.5%.

Finally, the quantum nature of QD emission was evaluated by performing intensity correlation measurements for the QD peak at 40K through the SiN grating coupler. The QD was pumped by the 633-nm CW laser with a power of 18 μW. Figure 4(d) shows a measured correlation histogram, exhibiting a clear antibunching dip at zero-time delay. Fitting to the curve yields a time-zero value of normalized second order coherence function $g^{(2)}(0) = 0.48$. This relatively low purity is primarily due to the background emission from the cavity mode, which is clearly visible in the spectrum in Fig. 4(a) and dominate 12% of the signal reaching the detector. Assuming that the background cavity emission is random noise in the intensity correlation measurement, the corrected $g^{(2)}(0)$ value yields 0.34 [32]. The unwanted cavity emission can be suppressed by resonant or phonon-assisted excitation of the QD.

In conclusion, we proposed and experimentally demonstrated a hybrid integration scheme for an InAs/GaAs QD-SPS on a SiN photonic integrated circuit using a coupled crossover waveguide structure. Using a 2D PhC nanocavity, the emission from a single QD was efficiently funneled into a GaAs waveguide and subsequently transferred to a single-mode SiN waveguide at the crossover structure. In this design, we numerically demonstrated 91.8% total coupling efficiency of QD radiation into the SiN waveguide. In experiment, we observed Purcell-enhanced single-photon emission, on-chip waveguiding, and outcoupling through a SiN grating coupler. The measured $g^{(2)}(0)$ was 0.48 (0.34) before (after) background noise correction and the estimated total coupling efficiency was 16.5%. The use of the coupled crossover waveguides unlocks the limitations in hybrid photonic integration of materials with largely different refractive indices. The 2D architecture offers profound design flexibility of the device and is advantageous for implementing various functions around the cavity region, such as electrodes and heaters. While the optical structure discussed in this study occupies a relatively large space on the chip, we envision that the footprint can be several times smaller by eliminating redundant parts.

**Acknowledgments**

We would like to thank T. Ito, Sh. Ito, S. Ji, H. Otsuki, M. Nishioka and S. Ishida for their technical support and fruitful discussions. The authors acknowledge support from grants-in-aid for scientific research KAKENHI (No. 22H01994, 22H00298, and 22H04962), JST FOREST Program (No. JPMJFR2113F), and JST SPRING Program (No. JPMJSP2123).

# Figures

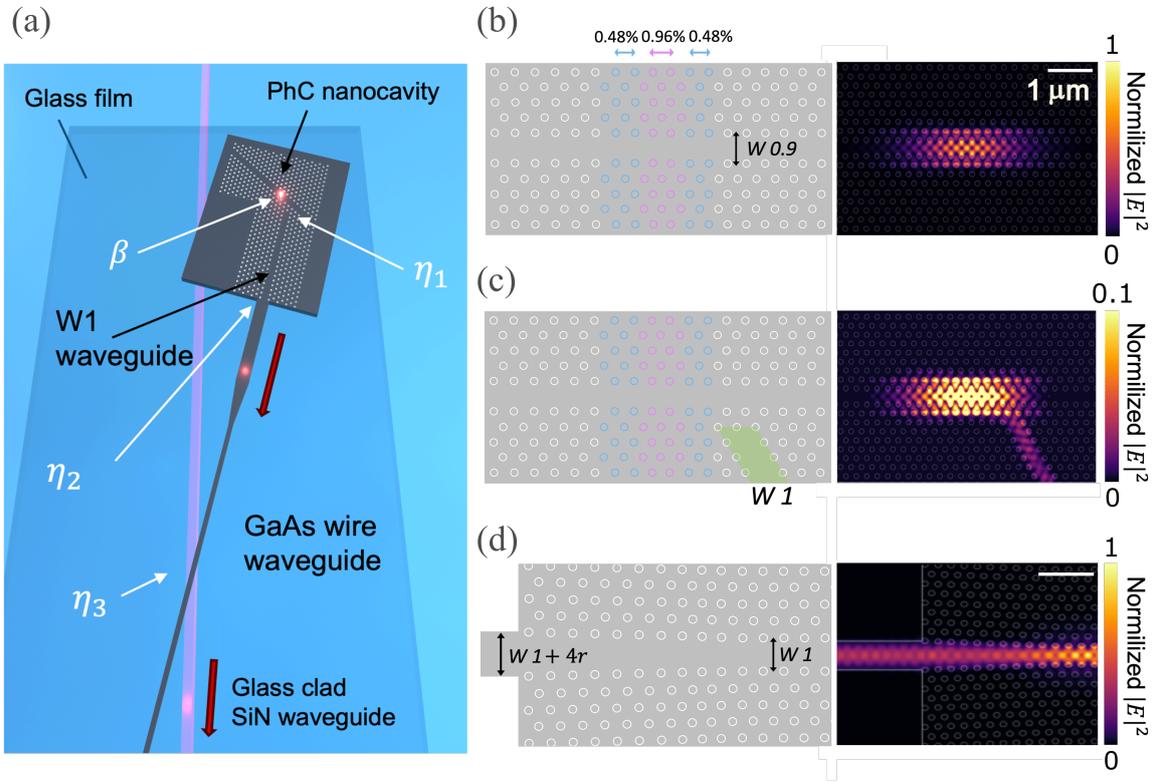

**Fig. 1.** (a) Schematic of the investigated SPS structure. (b) Detailed cavity structure and electric field distribution of the fundamental mode. (c) Same as in (b), but for the structure with a coupled W1 waveguide. (d) Adiabatic mode expander for connecting between the W1 and wire waveguides. The right panel shows an electric field distribution.

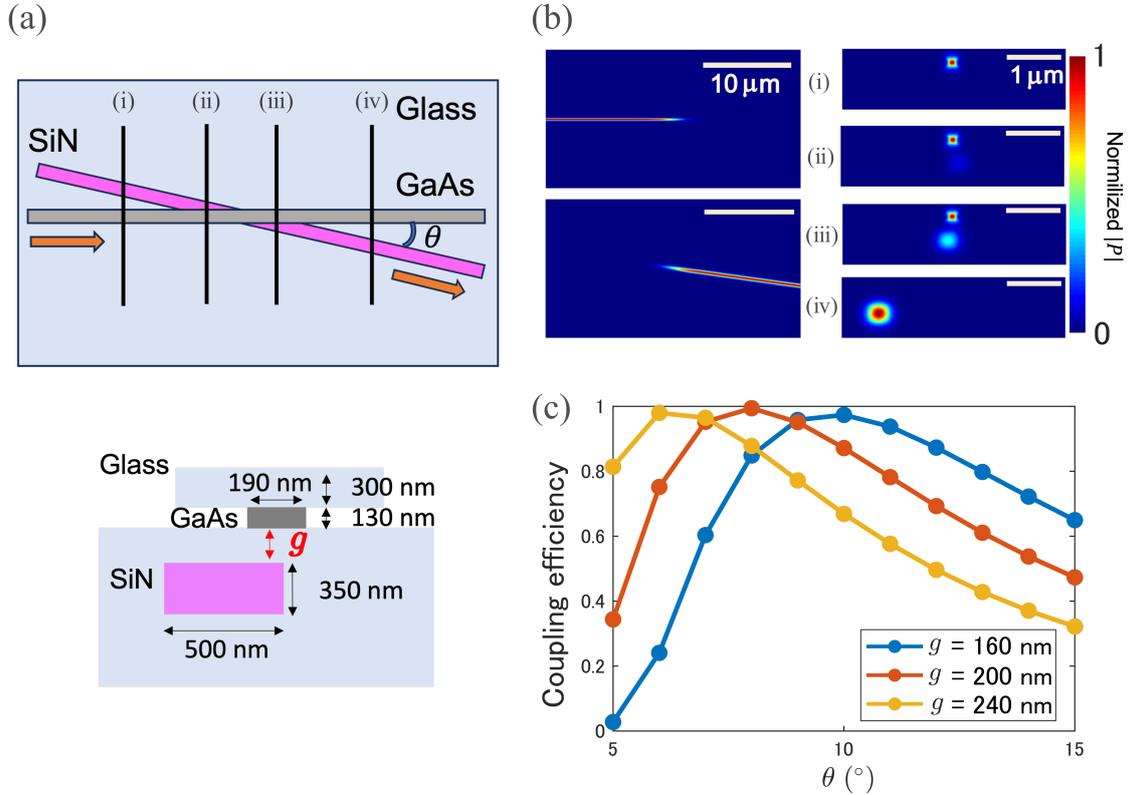

**Fig. 2.** (a) Coupled crossover waveguide structure. The top (bottom) panel shows a top (cross-sectional) view. (b) Optical power density distributions computed for the crossover waveguides at a crossing angle of 8° and $g$ = 200 nm. The left column shows top views recorded in the GaAs (upper) and SiN (lower) planes. The right column displays cross-sectional power distributions recorded in the slice positions shown in (a). (c) Simulated coupling efficiencies of the coupled crossover waveguide structure at 940 nm for different crossing angles and $g$.

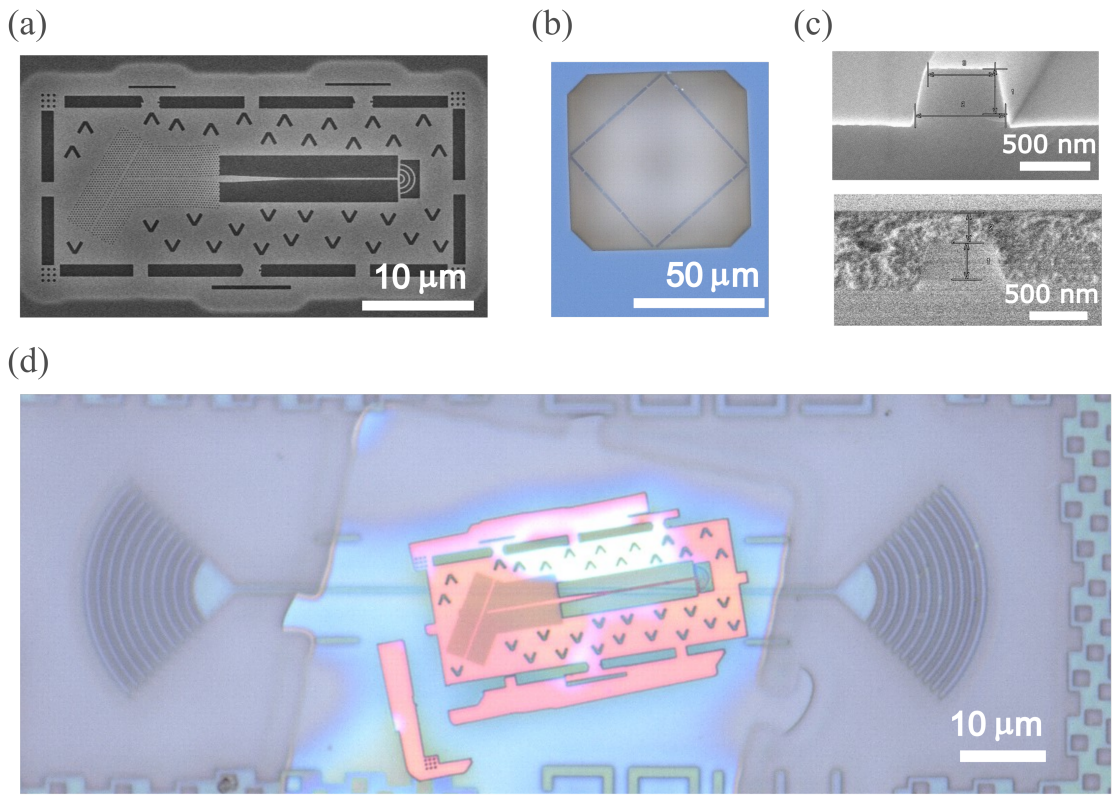

**Fig. 3.** (a) SEM image of a fabricated GaAs structure. (b) Microscope image of an air-suspended glass thin film. (c) Cross-sectional SEM image of a SiN waveguide (top) with and (bottom) without the clad. (d) Microscope image of a completed structure.

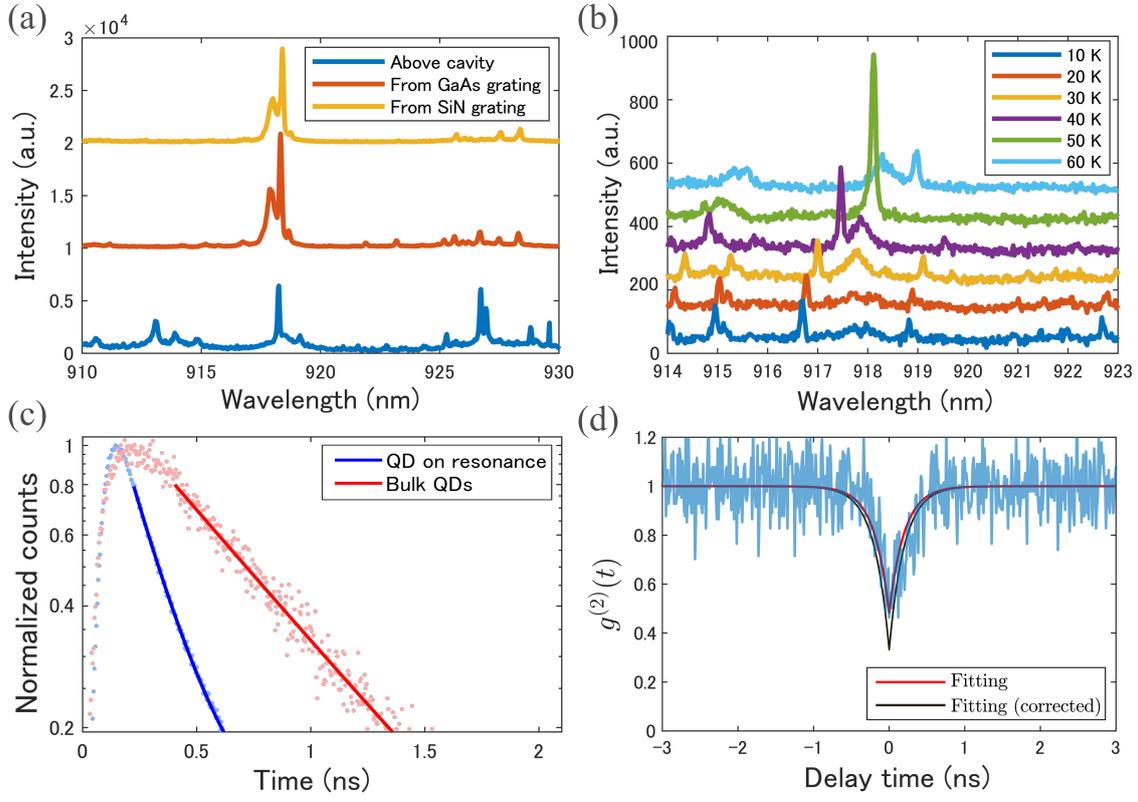

**Fig. 4.** (a) Measured PL spectra collected through the SiN or GaAs out-coupler, or above the cavity. (b) Temperature-dependent PL spectra recorded through the SiN grating. (c) Time-resolved PL signal of the QD in resonance with the cavity, plotted together with that of bulk QDs at 4 K. (d) Intensity correlation histogram measured for the QD peak at 40 K.